\documentclass[12pt]{article}
\usepackage{amssymb}
\usepackage[dvips]{graphicx}
\usepackage[dvips]{epsfig}
\usepackage{subfigure}

\newcommand{\beq}{\begin{equation}}
\newcommand{\eeq}{\end{equation}}
\newcommand{\bea}{\begin{eqnarray}}
\newcommand{\eea}{\end{eqnarray}}
\newcommand{\bn}{\begin{eqnarray}}
\newcommand{\en}{\end{eqnarray}}
\newcommand{\lam}{\lambda}
\newcommand{\vf}{\varphi}

\newcommand{\nn}{\nonumber}

\newcommand{\no}{\noindent}

\def\beq{\begin{eqnarray}}    
\def\eeq{\end{eqnarray}}      
\begin{document}

\title{\textbf{Unusual Yang-Lee edge singularity in the one-dimensional axial-next-to-nearest-neighbor Ising model.}}
\author{D. Dalmazi\footnote{dalmazi@feg.unesp.br}  and F. L. S\'a\footnote{ferlopessa@yahoo.com.br} \\
\textit{UNESP - Univ Estadual Paulista - Campus de Guaratinguet\'a - DFQ } \\
\textit{Av. Dr. Ariberto P. da Cunha, 333} \\
\textit{CEP 12516-410 - Guaratinguet\'a - SP - Brazil. }}
\maketitle

\begin{abstract}

We show here for the one-dimensional spin-$1/2$ ANNNI
(axial-next-to-nearest-neighbor-Ising) model in an external
magnetic field that the linear density of Yang-Lee zeros may
diverge with critical exponent $\sigma = -2/3$ at the Yang-Lee
edge singularity. The necessary condition for this unusual
behavior is the triple degeneracy of the transfer matrix
eigenvalues. If this condition is absent we have the usual value
$\sigma = -1/2$. Analogous results have been found in the
literature in the spin-1 Blume-Emery-Griffths model and in the
three-state Potts model in a magnetic field with two complex
components. Our results support the universality of  $\sigma =
-2/3$ which might be a one-dimensional footprint of a tricritical
version of the Yang-Lee-Edge singularity possibly present also in
higher-dimensional spin models.

\end{abstract}

\vfill\eject

\section{Introduction}

 In spin-$1/2$ models in an
external magnetic field $H$, the partition function  for finite
number of spins is proportional to a polynomial on the variable
$u=\exp(-2H/kT)$. Since polynomials are basically specified by
their zeros, those zeros furnish all relevant physical
information. Given that the partition function is a sum of
exponentials (positive numbers) with positive coefficients, it is
clear that we can only have zeros in the u-variable for complex
magnetic fields. Such zeros for complex magnetic fields are called
Yang-Lee zeros (YLZ) and are naturally studied on the complex-u
plane as we do here. Their relevance in the study of phase
transitions has been pointed out in  1952 by C. N. Yang and T. D.
Lee, see \cite{yl}.

In several spin models the YLZ tend, in the thermodynamic limit,
to form continuous curves on the complex u-plane. In particular,
it has been rigourously proved in \cite{ly} that all YLZ of the
spin-$1/2$ Ising model, even before the thermodynamic limit, lie
on the unit circle $\vert u \vert = 1$. This circle theorem has
been generalized, for example, to higher-spins Ising models
\cite{g69} and to include other interacting terms \cite{suzuki}.
As a rule, the zeros tend to leave the unit circle as the
ferromagnetic (FM) Ising coupling $J$ becomes smaller as compared
to other couplings. For more references see the review work
\cite{bena}. At $T \le T_c$ the YLZ tend to pinch the positive
real axis on the complex-$u$ plane at the first-order and
second-order phase transition points as we approach the
thermodynamic limit. However, if $T> T_c$ they accumulate at the
endpoints of the edges ($u_E=\exp(-2\beta H_E)$) of the curves
with a divergent linear density $\rho(u)\sim|u-u_E|^\sigma$ with
$\sigma < 0$.

As first noticed in \cite{kg} for the two-dimensional Ising model,
the power-like behavior is independent of the temperature as long
as $T > T_c$. The universality of the exponent $\sigma$ was
explained in \cite{fisher78} as a result of a usual critical point
described by a field theory with $i\, \Phi^3$ interaction vertex.
The corresponding endpoints  $u_E$ have been called Yang-Lee edge
singularities (YLES). In $D=2$ dimensions, the use of conformal
field theory \cite{cardy} predicts $\sigma=-1/6$. This result has
been verified for the Ising Model numerically \cite{ms,kim2006}
and also experimentally from magnetization data \cite{binek}.

According to arguments given in \cite{fisherp} one should have
$\sigma=-1/2$ in one dimension. Notwithstanding, even in $D=1$ the
exact position and density of the YLZ is not known analytically.
One exception is the one-dimensional spin-$1/2$ Ising model. In
this case the linear density was already known exactly \cite{ly}
furnishing $\sigma=-1/2$.  Numerical works have confirmed
$\sigma=-1/2$ in several one-dimensional spin models
\cite{kurze,uzelac, wang,ananikian} including, see \cite{hga}, the
same model discussed here. An early exception is a special type of
three-state Potts model \cite{mittag}. In such model the spins are
coupled with a magnetic field with two complex components. By
fine-tuning the couplings of the model it has been obtained
another type of YLES with $\sigma=-2/3$. More recently, inspired
by the work of \cite{ananikian} we \cite{ds1,ds2} have shown that
$\sigma=-2/3$ is not a peculiar feature of the special Potts model
used in \cite{mittag} but it is also present in the more familiar
spin-1 Blume-Capel \cite{bc} and spin-1 Blume-Emery-Griffths (BEG)
\cite{beg} models. Once again a fine-tuning of the couplings is
required, which is typical of a tricritical phenomenon. In
particular, we need to have a triple degeneracy of the transfer
matrix eigenvalues. So far one has found this unusual value for
$\sigma$ only in spin models with three states per site. Here we
investigate the one-dimensional spin 1/2
axial-next-to-nearest-neighbor Ising (ANNNI) model, and confirm
that the same unusual critical behavior also appears in models
with two states per site and next-to-nearest-neighbor interaction.
Our setup is based on the transfer matrix solution and the use of
finite size scaling relations which are shown to be satisfied by
the YLZ close to the unusual YLES.

\section{The one-dimensional ANNNI model}

The ANNNI model was introduced in \cite{elliot}, see also
\cite{fselke}, as a simple model to describe spatially modulated
periodic structures observed in magnetic and ferroelectric
materials. In its higher-dimensional $(D>1)$ versions the model
has interesting physical applications, see \cite{selkereview} for
a review work. We concentrate here in its one-dimensional version
as a simpler laboratory to investigate the existence of other
types of Yang-Lee-Edge singularities. The energy and the partition
function of the spin $1/2$ one-dimensional ANNNI model in an
external magnetic field are given by:

\beq E=-J\sum_i S_iS_{i+1}-K \sum_iS_iS_{i+2}-H\sum_iS_i \eeq \beq
Z=\sum_{\{S_i\}}e^{-\beta E} \eeq

\no where $S_i=\pm 1$, $J$ and $K$ are coupling constants between
nearest and next to nearest neighbor spins respectively and $H$ is
the magnetic field. Usually the model is defined with $J>0$
(ferromagnetic or FM) and $K<0$ (anti-ferromagnetic or AFM)
couplings. Here we start from a more general standpoint where $J$
and $K$ are arbitrary real numbers and the magnetic field $H$ can
assume complex values. We use the notation:

\beq c=e^{-\beta J} ,\quad d=e^{\beta K} , \quad u=e^{2\beta
H},\quad A=u+1/u = 2\, \cosh (2 \beta H). \label{eq:notacao} \eeq

The temperature is given in units of the nearest neighbor
coupling, i.e., the range $0\leq T<\infty$ corresponds to $0\leq
c\leq1$ if $J>0$ or $c \ge 1$ if $J<0$. Likewise $-\infty<K\leq 0$
implies $0\leq d \leq1$ while $ 0\le K < \infty$ leads to $d\ge
1$. Here we use periodic boundary conditions, $S_i=S_{i+N}$. The
partition function can be found via transfer matrix \cite{hga} :

\beq Z_N={\textrm{\normalsize Tr}}
\,\,T^N=\lam_1^N+\lam_2^N+\lam_3^N+\lam_4^N  \label{zn}\eeq

\no where the $\lam_1$, $\lam_2$, $\lam_3$ and $\lam_4$ are the
eigenvalues of the transfer matrix of the model and they can be
determined by the characteristic equation:

\beq \lam^4-a_3\lam^3+a_2 \lam^2-a_1\lam+a_0=0 \label{quartic}
\eeq

\no where

\beq
a_0&=&-\left(\frac{1}{d^2}-d^2\right)^2 \label{a0a}\\
a_1&=&d\, c\left(\frac{1}{d^2}-d^2\right)\sqrt{A+2} \label{a1a}\\
a_2&=&d^2\left(\frac{1}{c^2}-c^2\right)\label{a2a}\\
a_3&=& \frac{d}{c}\sqrt{A+2} \label{a3a} \eeq

\no  The ${\sf Z}_2$ symmetry $(H \to -H)$ of $Z_N$ is explicit in
the factor $\sqrt{A+2}=e^{\beta \, H} + e^{-\beta \, H}$ present
in $a_1$ and $a_3$. Notice that for $d=1$ we recover the
spin-$1/2$ Ising model with only two eigenvalues.

\section{$\sigma = -1/2$ and $\sigma = -2/3$ (Analytic approach)}

The partition function $Z_N$ is proportional to a polynomial of
degree $N$ in the ``fugacity'' $u$ (lattice gas interpretation).
All relevant information about $Z_N$ is contained in its zeros
$Z_N(u_k)=0,k=1,2,...,N$. Due to the ${\sf Z}_2$ symmetry one half
of the zeros is the inverse of the other half. For large number of
spins we assume, see \cite{nielsen,uzelac}, that the zeros are
determined by imposing that at least two of the eigenvalues of the
transfer matrix have the same absolute value which must be larger
than the other two remaining ones, i.e.,

\beq \lam_2=e^{i \varphi}\lam_1 \label{cd1}\eeq \beq
|\lam_2|=|\lam_1|>|\lam_i|, \quad i=3,4. \label{cd2}\eeq

\no For large $ N$ the contributions of $\lam_3$ and $\lam_4$ can
be disregarded and the partition function becomes

\beq Z_N\approx \lam_1^N \left(1+e^{i N \varphi}\right) \eeq

\no Therefore the zeros are given by

\beq \varphi_k= \left(2k-1\right)\frac{\pi}{N} ,\quad
k=1,2,..,N.\label{angles} \eeq

\no The equations (\ref{quartic}) and (\ref{cd1}) imply
\beq a_3&=&\lam_1\left(1+e^{i\varphi}\right)+\lam_3+\lam_4   \label{vieta1} \\
a_2&=& \lam_1^2e^{i\varphi}+\lam_1\left(\lam_3+\lam_4\right)\left(1+e^{i\varphi}\right)+\lam_3\lam_4   \label{vieta2}\\
a_1&=&\lam_1^2\left(\lam_3+\lam_4\right)e^{i\varphi}+\lam_1\lam_3\lam_4\left(1+e^{i\varphi}\right)   \label{vieta3} \\
a_0&=&\lam_1^2\lam_3\lam_4e^{i\varphi} \label{vieta4} \eeq

\no In (\ref{vieta1})-(\ref{vieta4}) we have four equations and
four variables $\lam_1 ,\lam_3 , \lam_4 , \vf $, eliminating
$\lam_1$, $\lam_3$ and $\lam_4$ we find an equation which depends
only on $\varphi$:

\beq
F(A,c,d,\varphi)&=&44\,a_0^3 -7\,a_1^4 + 34\,a_0 \,a_1^2\,a_2 - 28\,a_0^2\,a_2^2 - 2\,a_1^2\,a_2^3 + 6\,a_0\,a_2^4 - 40\,a_0^2\,a_1\,a_3 + 6\,a_1^3\,a_2\,a_3 \nonumber \\
&-& 24\,a_0\,a_1\,a_2^2\,a_3 + 34\,a_0^2\,a_2\,a_3^2 + a_1^2\,a_2^2\, a_3^2 - 2\,a_0\,a_2^3\,a_3^2 - 2\,a_1^3\,a_3^3 + 6\,a_0\,a_1\,a_2\,a_3^3  \nn \\
&-&  7\,a_0^2\,a_3^4+ 2\,\left\{ 40\,a_0^3 - a_0^2\,\left(
24\,a_2^2 + 34\,a_1\,a_3 - 29\,a_2\,a_3^2 + 6\,a_3^4 \right)
\right .\nn\\
&+&
\left .a_0\,\left[a_1^2\,\left( 29\,a_2 - 2\,a_3^2 \right) +a_2^3\,\left( 4\,a_2 - a_3^2 \right)+ a_1\,a_2\,a_3\,\left( -19\,a_2 + 5\,a_3^2 \right)  \right]\right .\nn\\
&-&\left . a_1^2\,\left[ 6\,a_1^2 + a_2^3 + a_1\,\left( -5\,a_2\,a_3+a_3^3 \right)\right]\right\}\,\cos (\varphi) + 2\,\left\{ 31\,a_0^3 + a_1^3\,\left( -3\,a_1 + a_2\,a_3 \right) \right .\nn\\
&+&\left.a_0\,\left[ a_2^4 + a_1^2\, \left( 18\,a_2 - a_3^2
\right)  +a_1\,a_2\,a_3\,\left( -8\,a_2+a_3^2 \right)\right] -
     a_0^2\,\left( 16\,a_2^2 - 18\,a_2\,a_3^2  \right.\right.\nn\\
&+&\left.\left.3\,\left( 8\,a_1\,a_3+a_3^4 \right)  \right]  \right\}\,\cos (2\varphi) + \left(40\,a_0^3- 2\,a_1^4\,+ 14\,a_0\,a_1^2\,a_2-16\,a_0^2\,a_2^2- 26\,a_0^2\, a_1\,a_3 \right. \nn\\
&-&\left. 2\,a_0\,a_1\,a_2^2\,a_3\, + 14\,a_0^2\,a_2\,a_3^2\,- 2\,a_0^2\,a_3^4\,\right)\cos (3\varphi) \nn\\
&+&\left( 20\,a_0^3 + 2\,a_0\,a_1^2\, a_2 - 4\,a_0^2\,a_2^2\,- 8\,a_0^2\,a_1\,a_3+2\,a_0^2\,a_2\,a_3^2\right)\cos(4\varphi) \nn\\
&+&\left(8\,a_0^3- 2\,a_0^2\,a_1\, a_3\right)\cos (5\varphi) +
2\,a_0^3\,\cos (6\varphi)=0 \label{F}\eeq

\no A similar\footnote{The corresponding expression printed in
\cite{hga}, differently from ours, does not lead to a double
degeneracy of the transfer matrix eigenvalues at $\vf =0$ as
expected.} expression has been derived before in \cite{hga} and
analogous formulae for the one-dimensional spin-1 Blume-Capel and
Blume-Emery-Griffiths models have appeared in \cite{ananikian} and
\cite{ds1} respectively\footnote{For the one-dimensional spin
$1/2$ Ising model the analogous of (\ref{F}) is simply
$A=2\sqrt{1-c^4}\, \cos\left(\varphi/2\right)$.}. We interpret
(\ref{F}) as a cubic equation for $A=A(\varphi)$ such that when we
plug it back in (\ref{quartic}) we get two eigenvalues with the
same absolute value according to (\ref{cd1}). Equation (\ref{F})
does not imply automatically the second condition (\ref{cd2}). In
practice we have to verify whether (\ref{cd2}) holds for each of
the three solutions $A_i(\varphi) \, , \, i=1,2,3$ of (\ref{F}).
At this point it is important to remark that, as its counterparts
in \cite{ananikian,hga,ds1}, the equation (\ref{F}) is symmetric
under $\varphi \to - \varphi$. This symmetry is not accidental. It
is a consequence of the permutation symmetry $\lam_1
\rightleftarrows \lam_2 $ hidden in (\ref{vieta1})-(\ref{vieta4}).
It becomes explicit when we perform $\left(\vf\, ,\, \lam_1\right)
\to \left(-\vf\, , \, e^{i \, \vf}\lam_1 \right)$. It is clear
that after eliminating $\lam_1 \, , \, \lam_3$ and $\lam_4$ the
resulting expression should be invariant under $\vf \to - \vf$.
Such symmetry will play a key role in determining the points in
the parameters space of the model where the unusual critical
behavior $\sigma=-2/3$ shows up. Next we show how (\ref{F}) can be
combined with a finite size scaling relation to determine the
exponent $\sigma$ analytically.

The closest zero $u_1[N]$ to the YLES $u_1[N\to\infty]$, for large
$N$, should obey \cite{Itzykson} the finite size scaling relation:

\beq \Delta \, u_1 (N) \equiv u_1[L] - u_1[\infty] =
\frac{C_1}{L^{y_h}}=\frac{C_1}{N^{y_h}} \quad . \label{fss1} \eeq

\no Where $y_h$ is the magnetic scaling exponent related to
$\sigma$ via $\sigma=(D-y_h)/y_h$. The constant $C_1$ is
independent on the number of the spins $N$ and $N=L^D=L$.

It is known \cite{Lassettre,Ashkin} that the degeneracy of the two
largest eigenvalues ($\lam_1=\lam_2$) of the transfer matrix
signalizes a second-order phase transition. Therefore, the YLES
occurs at $\varphi = 0$. Thus, for large $N$, we can assume that
the closest zero $u_1[N]$ is obtained from the smallest angle as
$u_1[N]=u[\varphi=\varphi_1=\pi/N]$, see (\ref{angles}). We obtain
$u(\varphi_1)$ from the appropriate solution $A(\varphi_1)$ of
(\ref{F}) via $u+1/u=A$. Expanding the result about $N\to\infty$,
which amounts to know $A(\varphi)$ in the vicinity of $\varphi=0$,
and comparing with (\ref{fss1}) we determine $y_h$ and $\sigma$
analytically.

Although the exact solutions of the cubic equation (\ref{F}) are
cumbersome, they can be used to show that an expansion of
$A(\varphi)$ about $\varphi=0$ contains only positive integer
powers of $\varphi $. The key ingredient is that the coefficients
of the cubic equation (\ref{F}) are analytic functions of $\vf$.
Therefore we can write down the large $N$ expansion

\bea A(\pi/N) &=&\left[A(0)+\varphi\left.\frac{dA}{d\varphi}\right|_{\varphi=0}+
\frac{\varphi^2}{2!}\left.\frac{d^2A}{d\varphi^2}\right|_{\varphi=0}+\frac{\varphi^3}{3!}\left.\frac{d^3A}{d\varphi^3}
\right|_{\varphi=0}+... \right]_{\varphi=\pi/N} \nn\\
&=&A(0)+\frac{\pi}{N}\left.\frac{dA}{d\varphi}\right|_{\varphi=0}+\frac{\pi^2}{2N^2}\left.\frac{d^2A}{d\varphi^2}
\right|_{\varphi=0}+\frac{\pi^3}{6N^3}\left.\frac{d^3A}{d\varphi^3}\right|_{\varphi=0}+...
\label{expansao A} \eea

\no Comparing (\ref{expansao A}) with (\ref{fss1}), if the lowest non-vanishing derivative at $\varphi=0$ is
$d^jA/d\varphi^j$ for some integer $j$ we have  $y_h = j$ and $\sigma=(1-j)/j$. In order to find $j$, instead of
using the complicated solutions of (\ref{F}), it is more elucidative to take consecutive derivatives of
(\ref{F}) with respect to $\varphi$. From the first derivative of (\ref{F}) we deduce, with help of
(\ref{vieta1})-(\ref{vieta4}), at $\varphi=0$ :

\beq
\left(\lam_1-\lam_3\right)^3\left(\lam_1-\lam_4\right)^3\left(\lam_3-\lam_4\right)^2\lam_1
\,d\left[c\left(\frac{1}{d^2}-d^2\right)+\frac{\lam_1^2}{c}\right]\left.\frac{dA}{d\varphi}\right|_{\varphi=0}=0
\quad . \label{1derivada} \eeq

\no Given that $\lam_1=\lam_2$ at $\varphi=0$, using
$\lam_1^2=c^2\left(d^2-\frac{1}{d^2}\right)$ in (\ref{quartic}) we
arrive at $\left(d^4-1\right)\left(c^4-1\right)=0$. Since $d$ and
$c$ are non negative numbers, see (\ref{eq:notacao}), and $c=1$
corresponds to $T\to\infty$ while $d = 1$ is the spin-$1/2$ Ising
model for which $\sigma=-1/2$ is known exactly , we assume
henceforth

\beq \lam_1^2 \ne c^2\left(d^2-\frac{1}{d^2}\right) \label{ineq}
\eeq

\no Equations (\ref{1derivada}) and (\ref{ineq}) tell us that

\beq \left.\frac{dA}{d\varphi}\right|_{\varphi=0}=0 \quad , \quad {\rm if} \,\, \lam_1=\lam_2\ne \lam_i \, , \,
i=3,4 \quad {\rm and} \quad \lam_3 \ne \lam_4 \label{dA1} \eeq

\no With help of (\ref{vieta1})-(\ref{vieta4}) and (\ref{dA1}),
the second derivative of (\ref{F}) at $\varphi=0$ furnishes:

\beq
\left(\lam_1-\lam_3\right)^3\left(\lam_1-\lam_4\right)^3\left(\lam_3-\lam_4\right)^2\lam_1\left\lbrace
2d\left[c\left(\frac{1}{d^2}-d^2\right)+\frac{\lam_1^2}{c}\right]
\left.\frac{d^2A}{d\varphi^2}\right|_{\varphi=0} \right.\nn \\
\left.
+\lam_1^2\left[(\lam_1-\lam_3)(\lam_1-\lam_4)\right]\right\rbrace=0
\label{2derivada} \eeq

\no From (\ref{ineq}),(\ref{dA1}) and (\ref{2derivada}) we have :

\beq \left.\frac{d^2A}{d\varphi^2}\right|_{\varphi=0}\not = 0
\quad , {\rm if} \,\, \lam_1=\lam_2\ne \lam_i \, , \, i=3,4 \quad
{\rm and} \quad \lam_3 \ne \lam_4 \label{dA2} \eeq

\no In summary, from (\ref{dA1}) and (\ref{dA2}) we conclude that
$y_h=2$ as long as we have neither a triple degeneracy of the
transfer matrix eigenvalues nor two double degeneracies.
Therefore, for the one-dimensional ANNNI model we show on general
grounds that $\sigma=-1/2$, see also \cite{hga}, except for the
two mentioned special cases where a different critical behavior
may appear in principle.

In the latter cases we have no information from (\ref{1derivada})
and (\ref{2derivada}). For triple degeneracy
($\lam_1=\lam_2=\lam_3$) the coefficients in (\ref{quartic}) must
satisfy:

 \bea
18\, a_1^2 - 14\, a_1a_2a_3 + 3\, a_1 a_3^3 -
a_2^2 a_3^2 + 4\, a_2^3 &=&    2\, a_0 \left(8\, a_2 - 3 a_3^2 \right)\label{tdcan1} \\
9\left\lbrack 12 \left(a_1^2-a_1a_2a_3\right) + 3\, a_1 a_3^3 -
a_2^2 a_3^2 \right\rbrack  +  32\, a_2^3  &=& 0 \label{tdcan2}
\eea

\no Working out (\ref{tdcan1}) and (\ref{tdcan2}) we arrive at :

\beq d^8(8 d^4 -9)(1-c^4)^2 + 108 \, c^4(1-d^4)^2 = 0 \quad ,
\label{tdc1} \eeq

\beq A = \frac{2(8\, d^4-7)}{8\, d^4 - 9} \label{tdc2} \eeq

\no Since both $c$ and $d$ are nonnegative real numbers, it is
clear from (\ref{tdc1}) that we can only have triple degeneracy if
$ 0 < d < (9/8)^{1/4}$. The condition (\ref{tdc1}) is a second
degree polynomial on $c^4$. Thus, there are only two possibilities
for the temperature as a function of $d$:

\beq c_{\pm}(d) = \left\lbrack \frac{\left(1-\frac 23 d^4\right)^{3/2} \pm (d^4-1)}{\left(1-\frac 23
d^4\right)^{3/2} \mp (d^4-1)}\right\rbrack^{1/4} \label{cpm} \eeq

\no Notice, in agreement with (\ref{tdc1}), that
$c_-(d)=1/c_+(d)$. In figure 1 we plot both $c_{\pm}(d^4)$. They
coalesce into $c_-=c_+=1$ ($T \to \infty$) at $d=1$. The function
$c_{+}(d)$ diverges at $d=\left(9/8\right)^{1/4}$ while $c_-(d)$
vanishes at that point. By inserting $c_{+}(d)$ in the exact
solutions of the cubic equation (\ref{F}) and expanding the
results about $\varphi=0$ we obtain:

\bea A_1 (\varphi) &=& \frac{ 2\,d^{12} - 40\,{d^8} + 117\,d^4 -81
}{d^{12}}-
  \frac{9\,\left(d^4 -1 \right) \,\left( 2\,d^4 -9 \right) \,\left( 2\,d^4 -3 \right) }{4\, d^{12}}\varphi^2 \nn\\
  &+&
  \frac{\left( d^4 -1 \right) \,\left\lbrace 3645 + 2\,d^4\,\left\lbrack -1944 + d^4\,\left( 297 + 64\,d^4 \right)
  \right\rbrack
  \right\rbrace
  }{216\, d^{12}}\varphi^4 + {\cal O}(\varphi^6)\label{A1} \eea

\bea A_2 (\varphi) &=& \frac{2\,\left( 8\, d^4 -7 \right) }{ 8\,
d^4 -9} + \frac{32 \, i\left( d^4 -1 \right) \,{\sqrt{ 2\,d^4 -3
}}}{ 9\left(8\,d^4 -9\right) }\varphi^3 -
  \frac{8\,\left( d^4 -1\right) }{27}\varphi^4 \nn\\ &-& \frac{2 \, i \left( d^4  -1 \right)
  \,\left( 8\,d^4  -9 \right)}{81{\sqrt{
  2\,d^4 -3 }}} \varphi^5 + {\cal O}(\varphi^6)\label{A2} \eea

\bea A_3 (\varphi) &=& \frac{2\,\left( 8\, d^4 -7 \right) }{ 8\,
d^4 -9} - \frac{32 \, i\left( d^4 -1 \right) \,{\sqrt{ 2\,d^4 -3
}}}{ 9\left(8\,d^4 -9\right) }\varphi^3 -
  \frac{8\,\left( d^4 -1\right) }{27}\varphi^4 \nn\\ &+& \frac{2 \, i \left( d^4  -1 \right)
  \,\left( 8\,d^4  -9 \right)}{81{\sqrt{
  2\,d^4 -3 }}} \varphi^5 + {\cal O}(\varphi^6)\label{A3} \eea


\no The solutions $A_2$ and $A_3$ are interchanged under the
symmetry $\varphi \to -\varphi$ of (\ref{F}) while $A_1$ is
invariant.

It turns out that only $A_2 (\varphi)$ and $A_3 (\varphi)$
satisfy, at $\varphi=0$, the triple degeneracy condition
(\ref{tdc2}). Indeed, substituting $c=c_+(d)$ and $A=A_i (\varphi)
$ in (\ref{quartic}) one obtains the four transfer matrix
eigenvalues $\lam_{\alpha}=\lam_{\alpha} (\varphi) \, , \,
\alpha=1,2,3,4$ for each solution $A=A_i (\varphi)\, , \,
i=1,2,3$. We have checked numerically for several values of $d$ in
the range $0 < d < (9/8)^{1/4}$ that $A=A_1(\varphi)$ only leads
to  double degeneracy $\lam_1 = \lam_2 $ at $\varphi=0$. Besides,
it is such that $\vert \lam_1 (\varphi)\vert =\vert \lam_2
(\varphi)\vert  < \vert \lam_i (\varphi)\vert  \, , \, i=3,4$ in
the neighborhood of $\varphi=0$. So, $A_1(0)$ does not correspond
to a true edge singularity. On the other hand, the function
$A_3(\varphi)$, though it leads to triple degeneracy
$\lam_1=\lam_2=\lam_3$ at $\varphi=0$, it is such that $\vert
\lam_1 (\varphi) \vert = \vert \lam_2 (\varphi) \vert $ is not the
largest absolute value in the vicinity of $\varphi=0$. Thus, we do
not have partition function zeros for $A=A_3(\varphi)$. For
$A=A_2(\varphi)$ we have checked that $\lam_1=\lam_2=\lam_3$ at
$\varphi=0$ and more importantly $\vert \lam_1 (\varphi) \vert =
\vert \lam_2 (\varphi) \vert > \vert \lam_i (\varphi)\vert \, , \,
i=3,4 \,  $ which confirms that we do have Yang-Lee zeros
approaching the YLES $A_2(0)$ for $A=A_2(\varphi)$.

From the above discussion and (\ref{A2}) we conclude that for the
fine-tuning $c=c_+(d)$ we assure a different critical behavior
with $j=3=y_h$ ($\sigma=-2/3$) for the density of Yang-Lee zeros
at the YLES.

Regarding the second possibility $c=c_-(d)=1/c_+(d)$ it is
possible to show that in this case we have the usual result
$j=2=y_h$ ($\sigma=-1/2$). Indeed, it can be checked analytically
that the cubic equation (\ref{F}) is symmetric under $c \to 1/c$.
However, there is no such symmetry in (\ref{quartic}). If we plug
$c=c_-(d)$ and $A_1 (\varphi)$ in (\ref{quartic}), it turns out
that $\lam_1 = \lam_2 $ at $\varphi=0$ and  $\vert \lam_1
(\varphi)\vert =\vert \lam_2 (\varphi)\vert  > \vert \lam_i
(\varphi)\vert  \, , \, i=3,4$ in the neighborhood of $\varphi=0$.
So we have a true YLES with, see (\ref{A1}), $j=2=y_h$
($\sigma=-1/2$). The other functions $A_2 (\varphi)$ and $A_3
(\varphi)$ lead to $\lam_1=\lam_2=\lam_3$ at $\varphi=0$ but
$\vert \lam_1 (\varphi) \vert = \vert \lam_2 (\varphi) \vert $ is
not the largest absolute value about $\varphi=0$. So, we do not
have Yang-Lee zeros in those cases. Analogously, the case of two
double degeneracies $\lam_1=\lam_2 \ne \lam_3=\lam_4$ leads only
to $\sigma = -1/2$.

We see from (\ref{tdc2}), which gives the location of the YLES in
the triply degenerated case, that we can only have $\sigma = -2/3$
either for zeros lying on the unit circle ($-2\le A \le 2$), with
$0 \le d \le 1 $ (AFM coupling $K<0$), or on the negative real
axis ($ A < -2$) which requires $ 1 < d < (9/8)^{1/4}$ (FM
coupling $K>0$). In the first case $ 0 \le c_+(d) \le 1$ (FM
coupling $J>0$) while in the second one $ c
> 1$ (AFM coupling
$J<0$). Therefore, the unusual critical behavior $\sigma =-2/3$
only occurs for couplings $J$ and $K$ of opposite nature.

As a final remark we note from (\ref{A2}) that  $j \ne 3$ if $d^4
= 3/2$, which coincides precisely, using $c=c_+(d)$, with the
quadruple degeneracy of the eigenvalues
$\lam_1=\lam_2=\lam_3=\lam_4$. However, in this case $ c_+$
becomes complex and it will be neglected here. Anyway, this is an
indication that different values for $\sigma$ are associated with
multiple degeneracies of the transfer matrix eigenvalues.

In  the next section our analytic results are confirmed by
numerical calculations of the Yang-Lee zeros.

\section{Numerical results}

Comparing the FSS relation (\ref{fss1}) for two rings of sizes
$N_a$ and $N_{a+1}$ we derive a numerical estimate for $y_h$

\beq
y_h=-\left[\ln\frac{N_{a+1}}{N_a}\right]^{-1}\ln\left[\frac{\Delta
u_1(N_{a+1})}{\Delta u_1(N_a)}\right] \quad .\label{yhe} \eeq

\no Where  either the imaginary or the real part of $\Delta
u_1(N)$ can be used. In the case of triple degeneracy, the YLES
$u_1^{\pm}(\infty)=u_E^{\pm}$ are known exactly by inverting the
relation $u_E + 1/u_E = A_E$ where $A_E$ is given in (\ref{tdc2})
for each value of $d$. Since $A_E \in \Re $ it follows that
$u_E^-=(u_E^+)^*$. So we choose $u_E=u_E^+$ without loss of
generality.

We also consider another finite size scaling relation
\cite{Itzykson,Creswick} for the linear density of zeros close to
the YLES:

 \beq
\rho(L)=C_2 \, L^{y_h-D} = C_2 N^{y_h-1} \label{fss2} \eeq

\no where  $C_2$ is a constant independent on the number of the
spins $N$ while $N=L^D=L$. Analogous to (\ref{yhe}) we can derive
from (\ref{fss2}):

\beq
y_h=1+\left[\ln\frac{N_{a+1}}{N_a}\right]^{-1}\ln\left[\frac{\rho(N_{a+1})}{\rho(N_a)}\right]
\label{yhrho} \eeq

\no The scaling exponents obtained from (\ref{yhe}) and
(\ref{yhrho}) will be called respectively $y_h^E$ and
$y_h^{\rho}$. More specifically, we use $y_h^{E,Re}$ and
$y_h^{E,Im}$ according to the use of real or imaginary parts of
$\Delta \, u_1(N)$. We stress that (\ref{yhe}) and (\ref{yhrho})
furnish independent numerical estimates for $y_h$, since
$\rho(N)=1/\left( N\vert u_1-u_2 \vert \right) $ depends upon the
first and second closest zeros, $u_1\, ,\,u_2$ to the YLES while
$y_h^E$ depends only upon the first one. Later on, we will
extrapolate the finite size results (\ref{yhe}) and (\ref{yhrho})
for $N\to\infty$ via BST (Burlish-Stoer) extrapolation algorithm
\cite{BulirschStoer,Schutz}.

The  partition function zeros  for a ring with $N$ sites (spins)
are obtained numerically with help of the software Mathematica
from an analytic expression for $Z_N$. Even for one-dimensional
spin models there are no analytic expressions for the Yang-Lee
zeros in general. In order to save computer time, instead of using
the analytic solution for $Z_N$  given in (\ref{zn}) in terms of
the transfer matrix eigenvalues or in terms of the trace of powers
of the transfer matrix as in ${\rm Tr}\, T^N$, we use an
alternative\footnote{The alternative formula (\ref{zn2}) has a
diagrammatic interpretation as a connected Feynman diagram of a
zero-dimensional Gaussian field theory \cite{ds2}.} exact
expression derived in \cite{ds2} for any spin model which can be
solved via a finite transfer matrix. Namely, since $\lam_i \, , \,
i=1,2,3,4$ are solutions of the secular equation $P_4(\lam) \equiv
\lam^4-a_3\lam^3+a_2 \lam^2-a_1\lam+a_0=0$, we have shown, formula
(11) of \cite{ds2}, that (\ref{zn}) can be identified with

\beq Z_N = - N \left\lbrace \ln \left\lbrack g^4 P_4 (1/g)
\right\rbrack\right\rbrace_{g^N} = -N\left[\ln\left(1-a_3g+a_2g^
2-a_1g^3+a_0g^4\right)\right]_{g^N} \quad , \label{zn2} \eeq

\no where $g$ is an arbitrary real variable (power counting
parameter)  and $\left\lbrack f(g) \right\rbrack_{g^N}$ stands for
the coefficient of the term of power $g^N$ in the Taylor series of
$f(g)$ about $g=0$. For the lowest powers $N=1,2,3$ the reader can
easily check, with help of (\ref{vieta1})-(\ref{vieta4}) at $\vf
=0$, that (\ref{zn2}) indeed reproduces the transfer matrix
solution (\ref{zn}).

 At each value of $d$, the expression
$c_+(d)$, see (\ref{cpm}), furnishes the corresponding  fine-tuned
temperature for triple degeneracy,. We can also invert $c_+(d)$
and obtain $d$ for each given value of $c$. If, for instance, we
choose $c=0.5$, the inversion of $c_+(d)$ leads to $d\approx
0.8680$. We have displayed in figures 2-4 the YLZ and half of the
corresponding YLES (with positive imaginary part) for $d\approx
0.8680$ and $c=0.48 \, , \, 0.50 \, , \, 0.52$. It turns out that
the triple degeneracy point is a turning point after which each
edge bifurcates into two new ones. Right above the triple
degeneracy point ($c>0.50$) we have checked (not shown here) that
at the endpoint of each of the two new edges, figure 4(b), the
critical exponent is the usual one $y_h=2$ ($\sigma = -1/2$) while
right before ($c < 0.50$) we have a crossover behavior flowing
from $y_h=2$ to $y_h=3$ as we approach $c=0.50$ from below.

At the triple degeneracy point we have  made a detailed analysis
of the scaling behavior of the zeros in the neighborhood of the
YLES which is located, see (\ref{tdc2}) at $d\approx 0.8680$, at
$A_E\approx 1.1031 $ ($u_E\approx 0.5516 \pm 0.8341\, i$). In this
case all Yang-Lee zeros lie on the unit circle, see figure 3.

The log-log fits in figures \ref{fig:5a} and \ref{fig:5b} confirm the FSS relations (\ref{fss1}) and
(\ref{fss2}). They furnish the estimates $y_h^{E}=2.9960$ (using the real pat of the zeros) and
$y_h^\rho=2.9866$.

In table (\ref{table:1}) we present the sequences $y_h(N_a)$
obtained from formulae (\ref{yhe}) and (\ref{yhrho}). In the last
line we have extrapolated our finite size results  $N\to\infty$ by
using the BST  \cite{BulirschStoer,Schutz} algorithm with
$\omega=1$. This algorithm approximates the original sequence
$y_h(N_a)$ by another sequence of ratios of polynomials with
faster convergence. The BST approach depends upon the real free
parameter $\omega$:
$y_h(N)=y_h(\infty)+\frac{A_1}{N^\omega}+\frac{A_2}{N^{2\omega}}+...$
 where $A_1,A_2,\cdots $ are $N$-independent constants. We plot the extrapolated quantity $y_h(\infty)$ for
$0.1 \leq\omega\leq 3.0$ altogether with their error bars in
figure 6. The error bar corresponds to twice the difference
between the values of $y_h$ obtained at the step before the last
one in the extrapolating sequence. In table (\ref{table:1}) we
have chosen $\omega=1$ because it provides a more stable result,
i.e., $\frac{d y_h}{d w}=0$. Clearly from table (\ref{table:1})
and figures 5 and 6 we have a result very close to $y_h=3$
($\sigma=-2/3$) at the triple degeneracy point. We have also
checked numerically other couples of values for $(c,d)$ satisfying
the triple degeneracy condition $c=c_+(d)$. The BST extrapolated
results for $y_h$ are very similar as well as their error bars.
Some caution is needed when the edges are nearly horizontal
(vertical) lines. In those cases the smallest error bars for
$y_h^E$ are obtained by the use of the real (imaginary) part of
the first zero respectively.

\section{Conclusion}

Usually, in one-dimensional spin models, the linear density of
partition function zeros diverges with a critical exponent $\sigma
= -1/2$ at edge singularities. The universality of $\sigma$ is
known for a long time \cite{kg,fisher78,fisherp} and checked
explicitly in $D=1$ in several models, see e.g.
\cite{kurze,uzelac,wang,ananikian,hga}. However, in the works
\cite{mittag,ds1,ds2} one has found another critical behavior
($\sigma = -2/3$). The models investigated in
\cite{mittag,ds1,ds2} have three-state per site and only
nearest-neighbor interaction. Here we have shown that $\sigma =
-2/3$ also appears in the one-dimensional spin-$1/2$ ANNNI model
which contains a next-to-nearest-neighbor interaction and only two
states per site. Our results support the universality of $\sigma =
-2/3$. As in \cite{ds1,ds2}, the triple degeneracy of the transfer
matrix (TM) eigenvalues is necessary to evade the well known
result $\sigma = -1/2$. Such condition requires a fine-tuning of
the couplings of the model which explains why the authors of
\cite{hga} have only found $\sigma=-1/2$ for the same model
treated here. So, rather than the number of states per site, the
important point is the dimension of the TM and the number of the
free parameters of the model to be fine-tuned.

The above argument signalizes that the same phenomenon might occur
in higher-dimensional spin models under special circumstances,
since for $D > 1$  the number of eigenvalues of the TM increases
with the size of the lattice. In particular, one could speculate
that this phenomenon might be behind the sudden drop from $\sigma
=-0.15(2)$ down to $\sigma=-0.365$ as reported in \cite{binek},
similarly to the drop from $\sigma=-1/2$ to $\sigma=-2/3$. In
\cite{binek} one obtains the linear density of Yang-Lee zeros for
the two-dimensional Ising model,  above the critical temperature,
indirectly from a function that fits the experimental
magnetization data from a sample of FeCl$_2$. The $2D$-Ising model
works as a prototype for  FeCl$_2$ in some temperature range. As
shown already in \cite{kg} the discontinuity of the magnetization
across the curve of zeros furnishes their density. So one has
indirect access to $\sigma$ experimentally.

Another interesting point is that the triple degeneracy condition
$c=c_+(d)$ as given in (\ref{cpm}) defines a transition point
between two different loci of Yang-Lee zeros. For $c < c_+(d)$ we
have an arc of the unit circle with two edges while for $c >
c_+(d)$ each edge bifurcates into two new edges with some fraction
of zeros leaving the unit circle, see figure 2 and figure 4. Our
figure 4 is similar to figure 2-c of \cite{hga}. We only have
$\sigma=-2/3$ at $c=c_+(d)$.

At last, we remark that the subtle breakdown of the permutation
symmetry between the two largest eigenvalues ($\lam_1
\rightleftarrows \lam_2$) is the key point in finding the unusual
critical behavior with $y_h=3 (\sigma=-2/3)$.

\section{Acknowledgments}

D.D. is partially supported by CNPq and F.L.S. is supported by
CAPES. A discussion with A. de Souza Dutra is gratefully
acknowledged.














\newpage

\begin{figure}
\begin{center}
\epsfig{figure=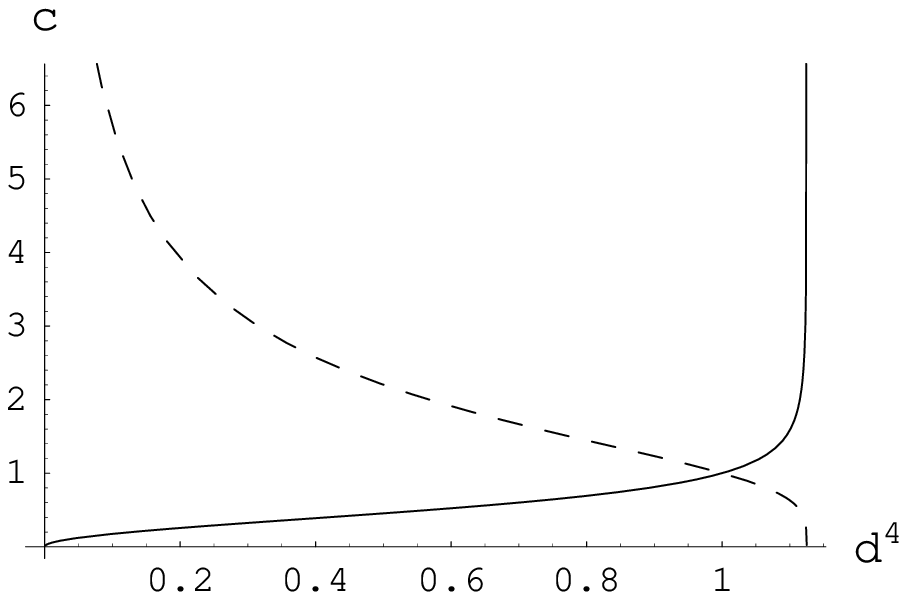,width=7cm,height=5cm} \caption{The
solid (dashed) line represents $c_+$ ($c_-$) as given in formula
(\ref{cpm}) }\label{fig:1}
\end{center}
\end{figure}

\begin{figure}[htb]
\center{\subfigure[ \label{fig:02a}]{\includegraphics[width=7cm,
height=7cm]{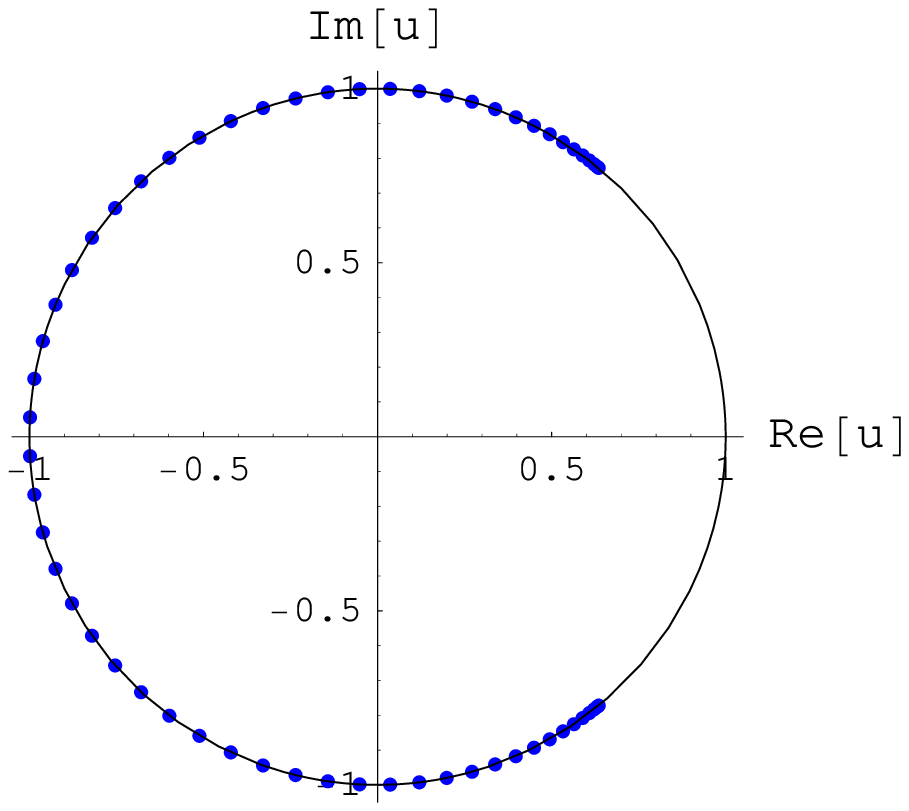}} \subfigure[
\label{fig:02b}]{\includegraphics[width=7cm,
height=7cm]{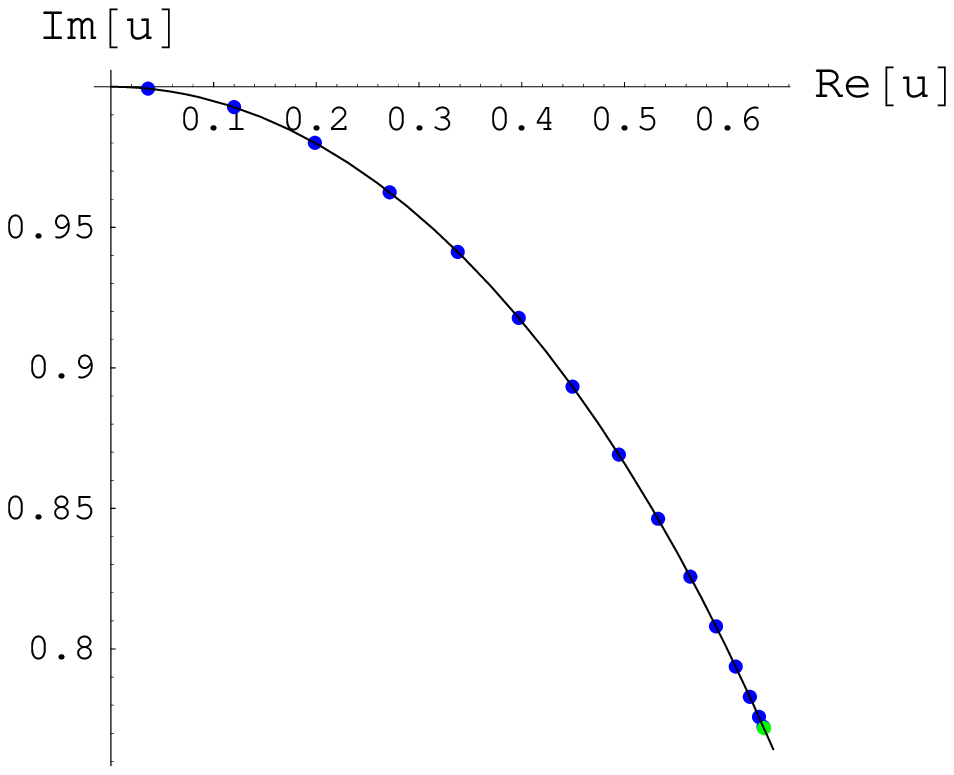}}} \caption{ (Color online) Yang-Lee
zeros for $N=60$ spins at $d\approx 0.8680$ and $c=0.48$. In
figure (a) we have all zeros while figure (b) displays only the
zeros on the first quadrant with the corresponding Yang-Lee edge
singularity (green (light gray) dot). The solid line stands for
$\vert u \vert = 1$ }\label{fig:02}
\end{figure}

\begin{figure}[htb]
\center{\subfigure[ \label{fig:03a}]{\includegraphics[width=7cm,
height=7cm]{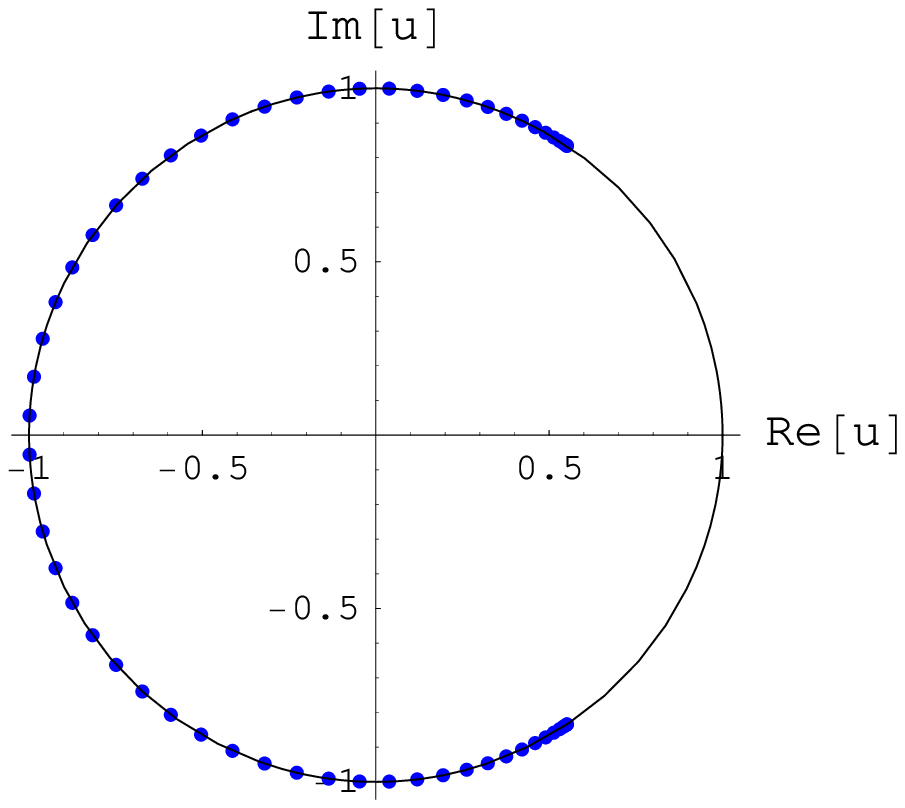}} \subfigure[
\label{fig:03b}]{\includegraphics[width=7cm,
height=7cm]{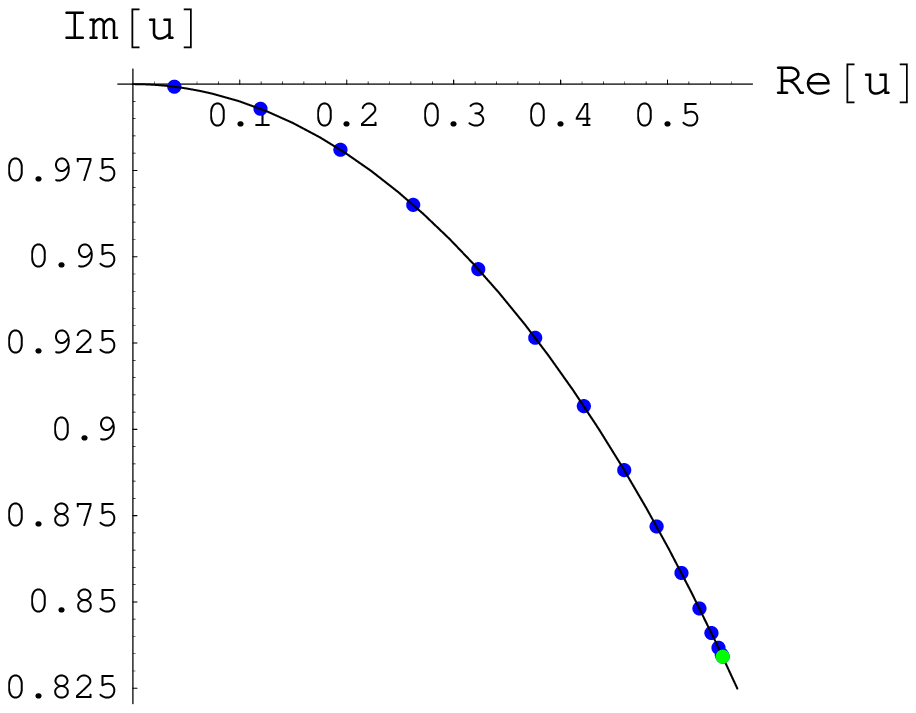}}} \caption{(Color online) Yang-Lee
zeros and half of the Yang-Lee edge singularities (green (light
gray) dot) for $N=60$ spins at $d\approx 0.8680$ and $c=0.5$
(triple degeneracy point)} \label{fig:03}
\end{figure}

\begin{figure}[htb]
\center{\subfigure[ \label{fig:04a}]{\includegraphics[width=7cm,
height=7cm]{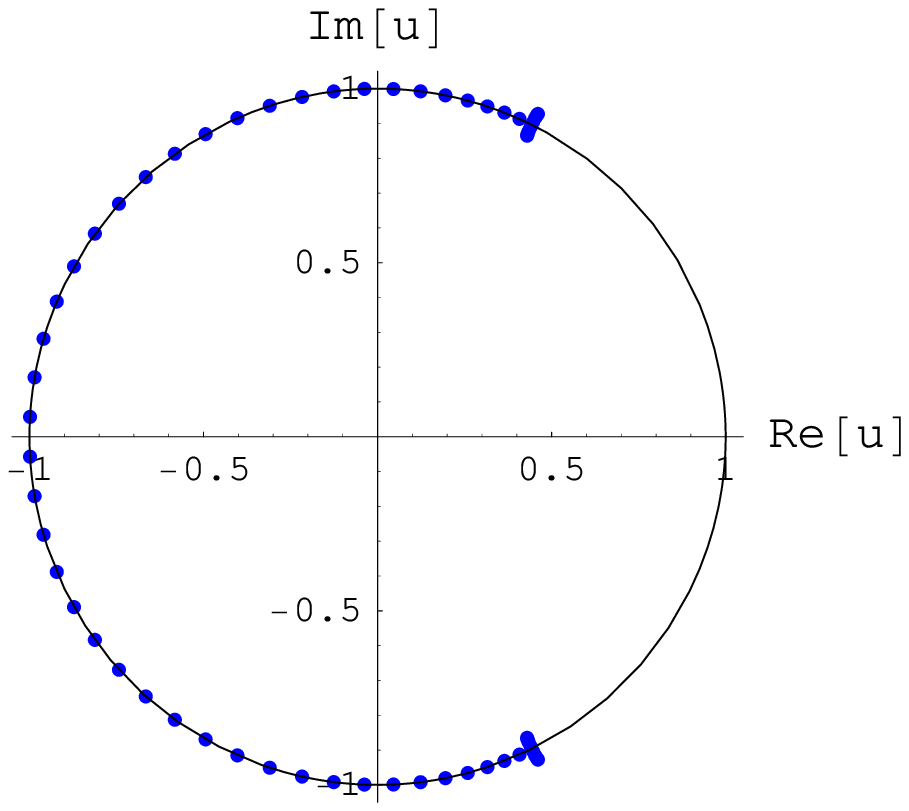}} \subfigure[
\label{fig:04b}]{\includegraphics[width=7cm,
height=7cm]{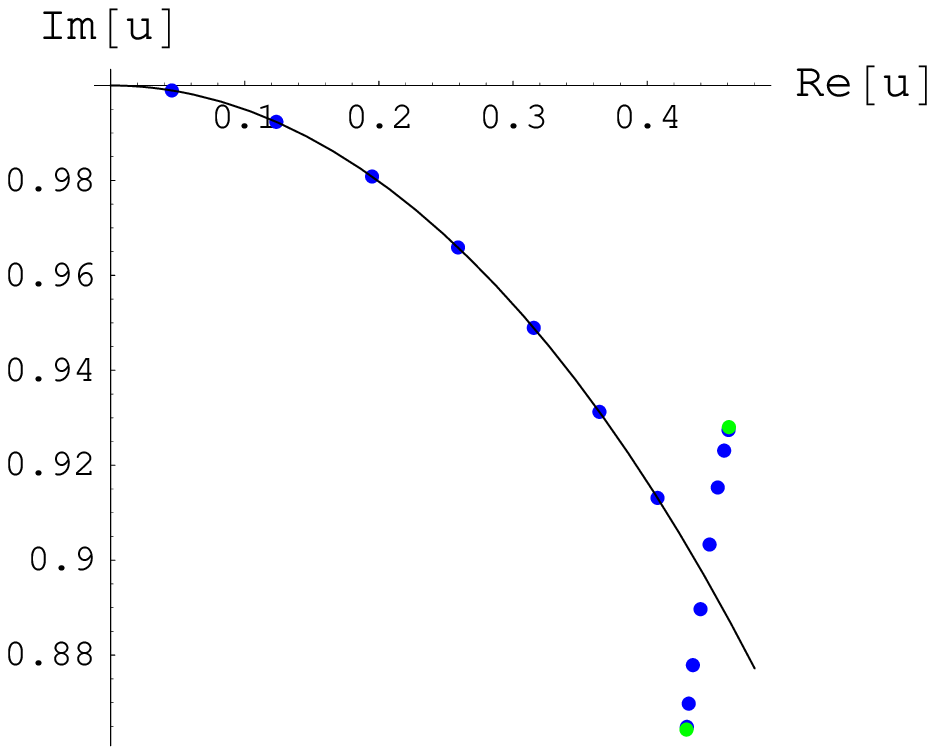}}} \caption{ (Color online) Yang-Lee
zeros and half of the Yang-Lee edge singularities (green (light
gray) dots) for $N=60$ spins at $d\approx 0.8680$ and $c=0.52$}
\label{fig:04}
\end{figure}

\begin{figure}[htb]
\center{\subfigure[Straight line: $\ln {\rm Re}(\Delta u) =
2.022-2.996\, \ln N$ with $\chi^2_E=9.30\times10^{-9}$
\label{fig:5a}]{\includegraphics[width=7cm,
height=6cm]{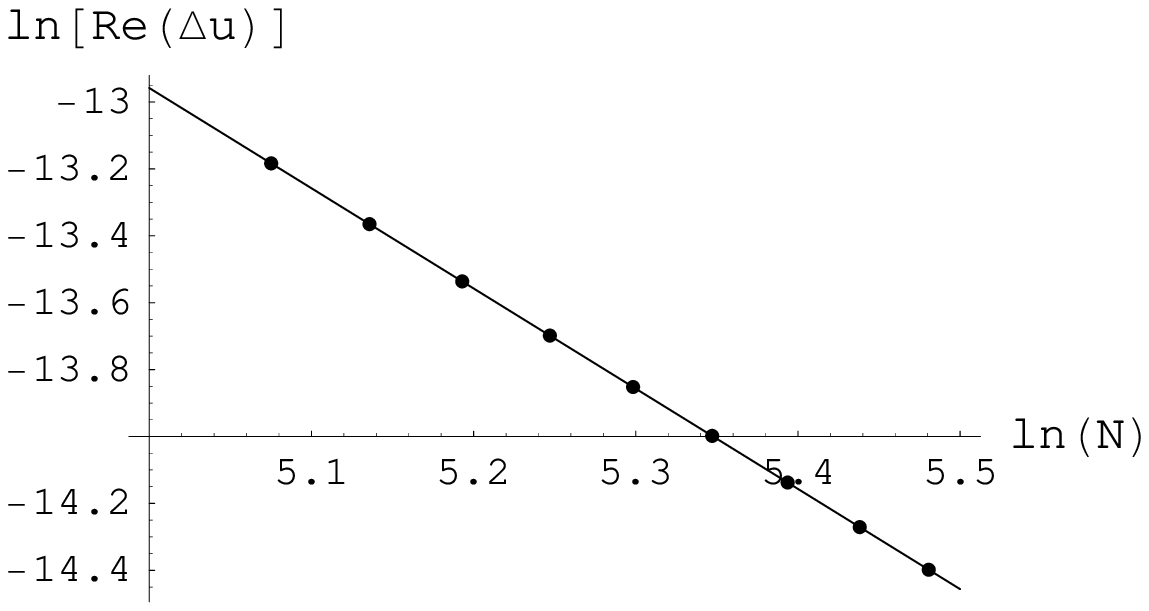}} \subfigure[Straight line: $\ln \rho=
-5.341+1.987\, \ln N$ with $\chi^2_\rho=9.03\times10^{-8}$
\label{fig:5b}]{\includegraphics[width=7cm,
height=6cm]{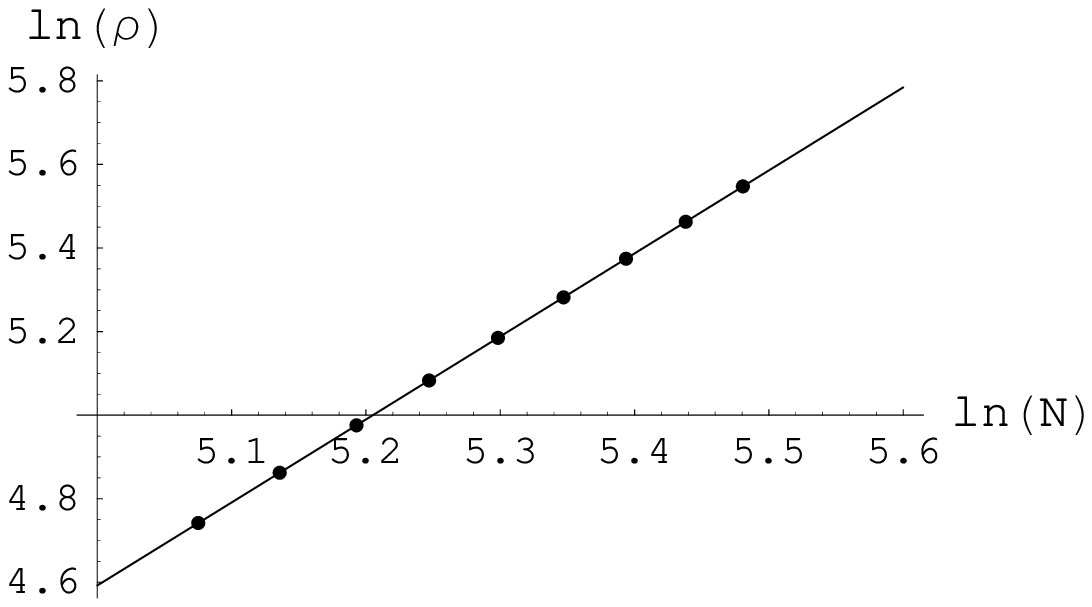}}} \caption{ Log-log fits of
(\ref{fss1}) and (\ref{fss2}) at $d\approx 0.8680$ and $c=0.5$
with $160 \le N \le 240$ spins}\label{fig:5}
\end{figure}

\begin{table}
\begin{center}
\begin{tabular}{|c|c|c|c|}
\hline
 $N_a$   & $ y_h^\rho$         &    $  y_h^{E,\,Re}$  & $y_h^{E,\,Im}$   \\
\hline
160      & 2.984155688412  & 2.995280492398    & 2.995273778849 \\
170      & 2.985057758429  & 2.995549136960    & 2.995543509484 \\
180      & 2.985862192536  & 2.995788825553    & 2.995784061898 \\
190      & 2.986584090647  & 2.996004003019    & 2.995999935011 \\
200      & 2.987235580790  & 2.996198248276    & 2.996194746764 \\
210      & 2.987826518204  & 2.996374475494    & 2.996371439968 \\
220      & 2.988364995418  & 2.996535081652    & 2.996532432944 \\
230      & 2.988857720521  & 2.996682056444    & 2.996679731514 \\
$\infty$ & 3.00000000000(1) & 3.00000000000(1)  & 3.00000000000(3)\\
\hline
\end{tabular}
\end{center}
\caption{Finite size results for the Yang-Lee zeros, where $
y_{h}^\rho$ are obtained from  (\ref{yhrho}) while
$y_{h}^{E,\,Re}$ and $y_{h}^{E,\,Im}$ come from (\ref{yhe}) using
the real and imaginary parts of the zeros respectively. The data
of this table have been obtained from rings with $160 \le N \le
240$ spins at $c=0.5$ and $d\approx 0.8680$. The last row is the
$N \to \infty$ extrapolation via BST algorithm with
$\omega=1$.}\label{table:1}
\end{table}

\begin{figure}[htb]
\center{\subfigure[ \label{fig:6a}]{\includegraphics[width=7cm,
height=5cm]{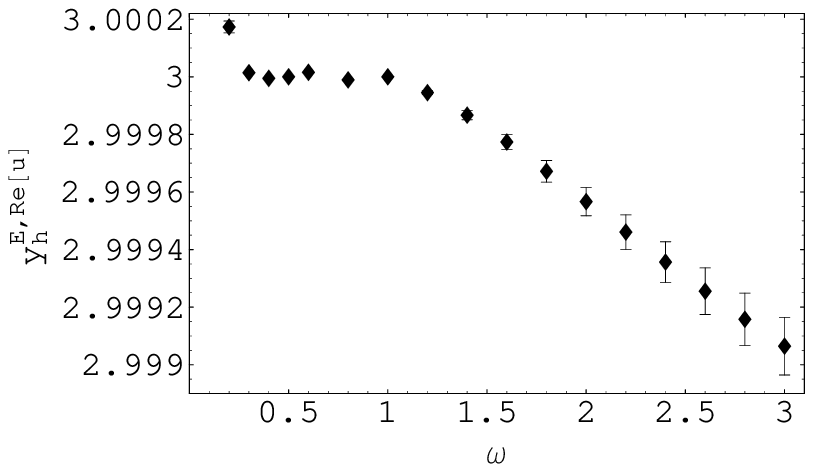}} \subfigure[
\label{fig:6b}]{\includegraphics[width=7cm,
height=5cm]{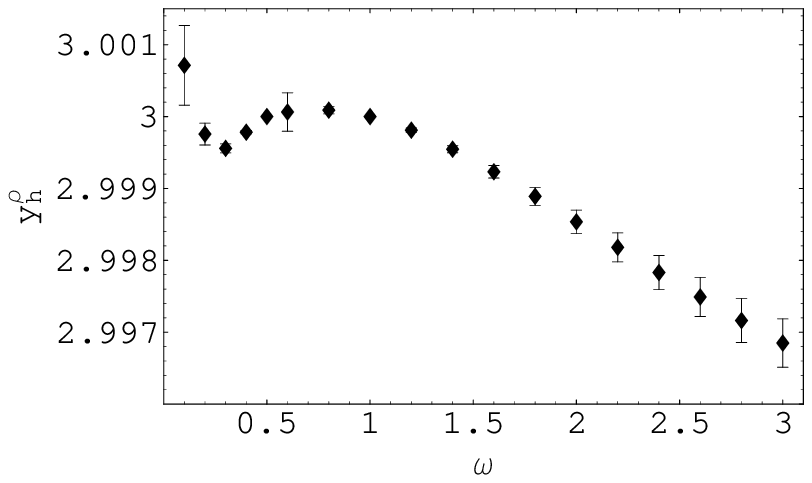}}} \caption{BST extrapolation for $0.1
\leq\omega\leq 3.0$ of $y_h^E$ using the real part of the zeros
(figure \ref{fig:6a}) and $y_h^{\rho}$ (figure \ref{fig:6b}) at
$d\approx 0.8680$ and $c=0.5$.}\label{fig:6}
\end{figure}

%
%
%
%
%
%
%

\end{document}